\documentclass[a4paper,11pt]{article} 

\usepackage[T1]{fontenc}
\usepackage[utf8]{inputenc} 
\usepackage[english]{babel} 

\usepackage{upgreek} 

\usepackage[colorlinks=true,urlcolor=blue]{hyperref}

\usepackage{lmodern} 

\usepackage{tabularx}
\usepackage{multirow}
\usepackage{multicol}

\usepackage{graphicx} 
\usepackage{caption} 
\usepackage{subcaption} 
\usepackage{wrapfig} 
\usepackage{here} 

\usepackage{float}
\usepackage{array}

\usepackage{tikz} 
\usepackage{tikz-cd}
\usetikzlibrary{arrows,calc,chains,matrix,positioning,scopes} 

\usepackage[top=3.cm, bottom=3.cm, left=2.5cm, right=2.5cm]{geometry} 

\usepackage{scalerel}
\usepackage{stackengine}

\usepackage{yhmath}
\usepackage{amsthm} 
\usepackage{amsmath} 
\usepackage{amsfonts}
\usepackage{amssymb} 
\usepackage{mathrsfs} 
\usepackage{esint} 
\usepackage{empheq}
\usepackage{stmaryrd}
\usepackage{blkarray} 

\usepackage{cancel} 

\DeclareSymbolFont{UPM}{U}{eur}{m}{n}
\DeclareMathSymbol{\uppartial}{0}{UPM}{"40}

\newcommand{\opd}{\mathop{}\mathopen{}\mathrm{d}}

\newcommand{\abs}[1]{\left\lvert#1\right\rvert}

\newcommand{\ensemblenombre}[1]{\mathbb{#1}}

\newcommand{\Z}{\ensemblenombre{Z}}

\newcommand{\R}{\ensemblenombre{R}}

\newcommand{\slfrac}[2]{\left.#1\middle/#2\right.}

\theoremstyle{definition}

\theoremstyle{plain}

\begin{document}

	\title{Exact computations \\ in \\ topological abelian Chern-Simons and BF theories}
	\date{\today}
	\author{Philippe \textsc{Mathieu} \\ {} \\ LAPTh, Université de Savoie, CNRS, Chemin de Bellevue, BP 110 \\  F-74941 Annecy-le-Vieux Cedex, France.}		
	\maketitle

\begin{abstract}

We introduce Deligne cohomology that classifies $U\left(1\right)$ fibre bundles over $3$-manifolds endowed with connections. We show how the structure of Deligne cohomology classes provides a way to perform exact (non-perturbative) computations in $U\left(1\right)$ Chern-Simons theory (resp. BF theory) at the level of functional integrals. The partition functions (and observables) of these theories are strongly related to topological invariants well-known by the mathematicians.

\end{abstract}
	
\section{Introduction}

Consider the following actions:
\begin{equation}
\label{Action_CS}
S_{\mathrm{CS}_{N}}\left[A\right] = 2\pi N\int_{\R^{3}}A\wedge dA = 2\pi N\int_{\R^{3}}A\wedge F
\end{equation}
and:
\begin{equation}
\label{Action_BF}
S_{\mathrm{BF}_{N}}\left[A,B\right] = 2\pi N\int_{\R^{3}}B\wedge dA = 2\pi N\int_{\R^{3}}B\wedge F
\end{equation}
where $A$ and $B$ are $U\left(1\right)$ connections. Here, the coupling constant $N$ is any real number.

The gauge transformation $A\longrightarrow A+d\Lambda$ where $\Lambda$ is a function leaves the actions \eqref{Action_CS} and \eqref{Action_BF} invariant. Since in the quantum context we consider the complex exponential of the action, the invariance required is less restrictive. Indeed, we can consider an invariance of $S$ up to an integer:
\begin{equation}
S \longrightarrow S + 2\pi n,\,\,n\in\Z\,\,\Rightarrow\,\, e^{iS} \longrightarrow e^{iS} 
\end{equation}
which implies that $N$ is quantized. Studying the gauge invariance properties of the holonomies, which are the observables of Chern-Simons and BF theories, it turns out that the most general gauge transformation is $A\longrightarrow A+\omega_{\Z}$ where $\omega_{Z}$ is a closed $1$-form with integral periods. On a contractible open set this transformation reduces to the classical one since, by Poincaré Lemma, there exists $\Lambda$ such that $\omega_{\Z}=d\Lambda$. In particular, this is the case when the theory is defined in $\R^{3}$ which is a contractible space. However, this generalized gauge transformation enables to define a theory on any closed (that is, compact without boundary) $3$-manifold $M$. The classical gauge transformation appears thus to be a particular case of the quantum one. 

In this paper we will consider the equivalence classes according to this quantum gauge transformation. These classes classify $U\left(1\right)$ fibre bundles over $M$ endowed with connections and their collection is the so-called first Deligne cohomology group of $M$. We will show that this structure enables to perform exact computations in the framework of $U\left(1\right)$ Chern-Simons and BF theories. 

\section{Deligne cohomology}

The most general statement we can start from is a collection of local gauge fields $A_{\alpha}$ in open sets $U_{\alpha}$ that cover the manifold $M$ we are considering. We suppose these open sets and their intersections to be contractible, so that we can in particular use the Poincaré Lemma inside. To define a global field, we need to explain how $A_{\alpha}$ and $A_{\beta}$ stick together in the intersection $U_{\alpha}\cap U_{\beta}$. This, by definition, is done thanks to a gauge transformation:
\begin{equation}
A_{\beta}=A_{\alpha}+d\Lambda_{\alpha\beta} \mbox{ in } U_{\alpha\beta}=U_{\alpha}\cap U_{\beta}
\end{equation}
The antisymetry of this relation in $\alpha$ and $\beta$ implies that $d\left(\Lambda_{\alpha\beta}+\Lambda_{\beta\gamma}+\Lambda_{\gamma\alpha}\right)=0$, making $\Lambda_{\alpha\beta}+\Lambda_{\beta\gamma}+\Lambda_{\gamma\alpha}$ a constant in $U_{\alpha}\cap U_{\beta}\cap U_{\gamma}$ that is an integer\footnote{since \eqref{Cocycle_Condition} is nothing but the cocycle condition for a $U\left(1\right)$ fibre bundle.}:
\begin{equation}
\label{Cocycle_Condition}
\Lambda_{\alpha\beta}+\Lambda_{\beta\gamma}+\Lambda_{\gamma\alpha}=n_{\alpha\beta\gamma}\in\Z \mbox{ in } U_{\alpha\beta\gamma}=U_{\alpha}\cap U_{\beta}\cap U_{\gamma}
\end{equation}
The symmetry in $\alpha$, $\beta$ and $\gamma$ of this last relation implies that:
\begin{equation}
n_{\alpha\beta\gamma}-n_{\alpha\beta\delta}+n_{\alpha\gamma\delta}-n_{\beta\gamma\delta}=0
\end{equation}

Thus, the generalization of our gauge potential on any closed $3$-manifold $M$ imposes to consider a collection $\left(A_{\alpha},\Lambda_{\alpha\beta},n_{\alpha\beta\gamma}\right)$ constituted of a family of potentials $A_{\alpha}$ defined in open sets $U_{\alpha}$, a family of functions $\Lambda_{\alpha\beta}$ defined in the double intersections $U_{\alpha\beta}$ and a family of integers defined in the triple intersections $U_{\alpha\beta\gamma}$ (all those open sets and intersections being contractible). Elements of those collections are related by:
\begin{equation}
\left\lbrace
\begin{array}{l}
A_{\beta} = A_{\alpha} + \opd\Lambda_{\alpha\beta} \mbox{ in } U_{\alpha\beta} \\
\Lambda_{\alpha\beta} + \Lambda_{\beta\gamma} + \Lambda_{\gamma\alpha} = n_{\alpha\beta\gamma}\in\Z \mbox{ in } U_{\alpha\beta\gamma}\\
n_{\alpha\beta\gamma} - n_{\alpha\beta\delta} + n_{\alpha\gamma\delta} - n_{\beta\gamma\delta} = 0
\end{array}
\right.
\end{equation}
These statements define a Deligne cocyle.

We need now to describe how this collection transforms when we perform a gauge transformation of the $A_{\alpha}$:
\begin{equation}
A_{\alpha} \longrightarrow A_{\alpha}+dq_{\alpha} \mbox{ in } U_{\alpha}
\end{equation}
where the family of $q_{\alpha}$ is a family of functions defined in the $U_{\alpha}$. This implies that the $\Lambda_{\alpha\beta}$ have to transform according to:
\begin{equation}
\Lambda_{\alpha\beta}\longrightarrow \Lambda_{\alpha\beta} + q_{\alpha} - q_{\beta} - m_{\alpha\beta} \mbox{ in } U_{\alpha\beta}
\end{equation}
where the family $m_{\alpha\beta}$ consists in integers, mainly because the $n_{\alpha\beta\gamma}$ are. Finally, the $n_{\alpha\beta\gamma}$ transform thus according to:
\begin{equation}
n_{\alpha\beta\gamma}\longrightarrow n_{\alpha\beta\gamma} - m_{\beta\gamma} + m_{\alpha\gamma} - m_{\alpha\beta} \mbox{ in } U_{\alpha\beta\gamma}
\end{equation}

Hence, the collection $\left(q_{\alpha},m_{\alpha\beta}\right)$ where $q_{\alpha}$ are functions defined in the $U_{\alpha}$ and $m_{\alpha\beta}$ are integers defined in the intersections $U_{\alpha\beta}$ together with the set of rules: 
\begin{equation}
\left\lbrace
\begin{array}{l}
A_{\alpha} \longrightarrow A_{\alpha}+dq_{\alpha} \mbox{ in } U_{\alpha} \\
\Lambda_{\alpha\beta}\longrightarrow \Lambda_{\alpha\beta} + q_{\alpha} - q_{\beta} - m_{\alpha\beta} \mbox{ in } U_{\alpha\beta}\\
n_{\alpha\beta\gamma}\longrightarrow n_{\alpha\beta\gamma} - m_{\beta\gamma} + m_{\alpha\gamma} - m_{\alpha\beta} \mbox{ in } U_{\alpha\beta\gamma}
\end{array}
\right.
\end{equation}
generalize the idea of gauge transformation. These rules define the addition of a Deligne coboundary to a Deligne cocycle. The quotient set of Deligne cocycles by Deligne coboundaries is the first Deligne cohomology group $H^{\left[1\right]}_{D}$.

\section{Structure of the space of Deligne cohomology classes}

$H^{\left[1\right]}_{D}$ is naturally endowed with a structure of $\Z$-modulus. It can be described in particular through two exact sequences. The first one is:
\begin{center}
\includegraphics[scale=1.]{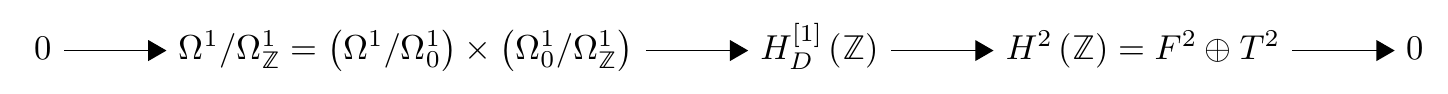}
\end{center}
where $\Omega^{1}/\Omega_{\Z}^{1}$ is the quotient of the $1$-forms by the closed $1$-forms with integal periods and  $H^{2}\left(\Z\right)$ is the space of cohomology classes of the manifold. This is an abelian group, which can thus be decomposed as a direct sum of a free part $F^{2} = \Z^{b^{2}}$ and a torsion part $T^{2}=\Z_{p_{1}}\oplus\cdots\oplus\Z_{p_{n}}$. This exact sequence shows that the space of Deligne cohomology classes can be thought as a set of fibres over the discrete net constituted by $H^{2}\left(\Z\right)$ and inside which we can move thanks to elements of $\Omega^{1}/\Omega_{\Z}^{1}$.  
\begin{center}
\includegraphics[scale=1.]{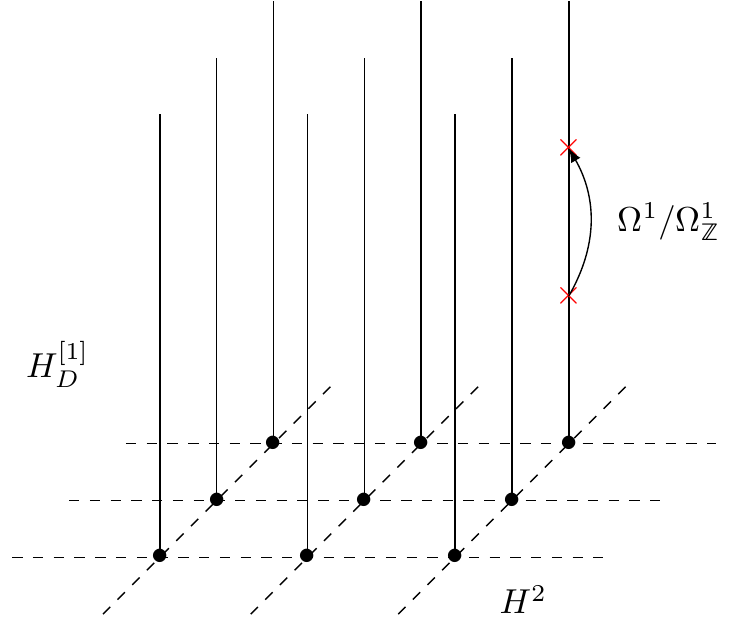}
\end{center}

The second exact sequence that enables to represent $H^{\left[1\right]}_{D}$ is:
\begin{center}
\includegraphics[scale=1.]{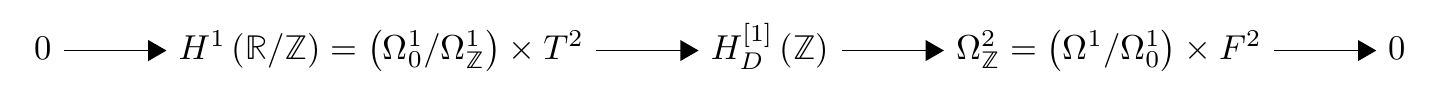}
\end{center}
where $H^{1}\left(\R/\Z\right)$ is the first cohomology group $\slfrac{\R}{\Z}$-valued and $\Omega^{2}_{\Z}$ is the set of closed $2$-forms with integral periods. This exact sequence leads to the following representation:
\begin{center}
\includegraphics[scale=1.]{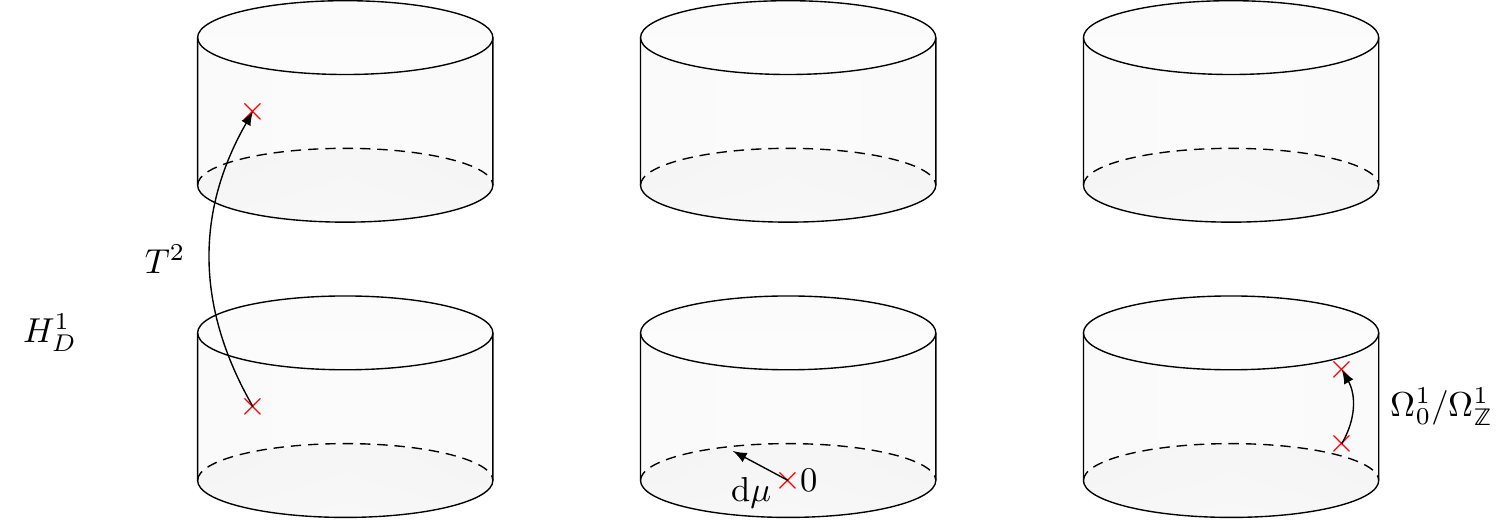}
\end{center}

Those two exact sequences contain the same information:
\begin{center}
\includegraphics{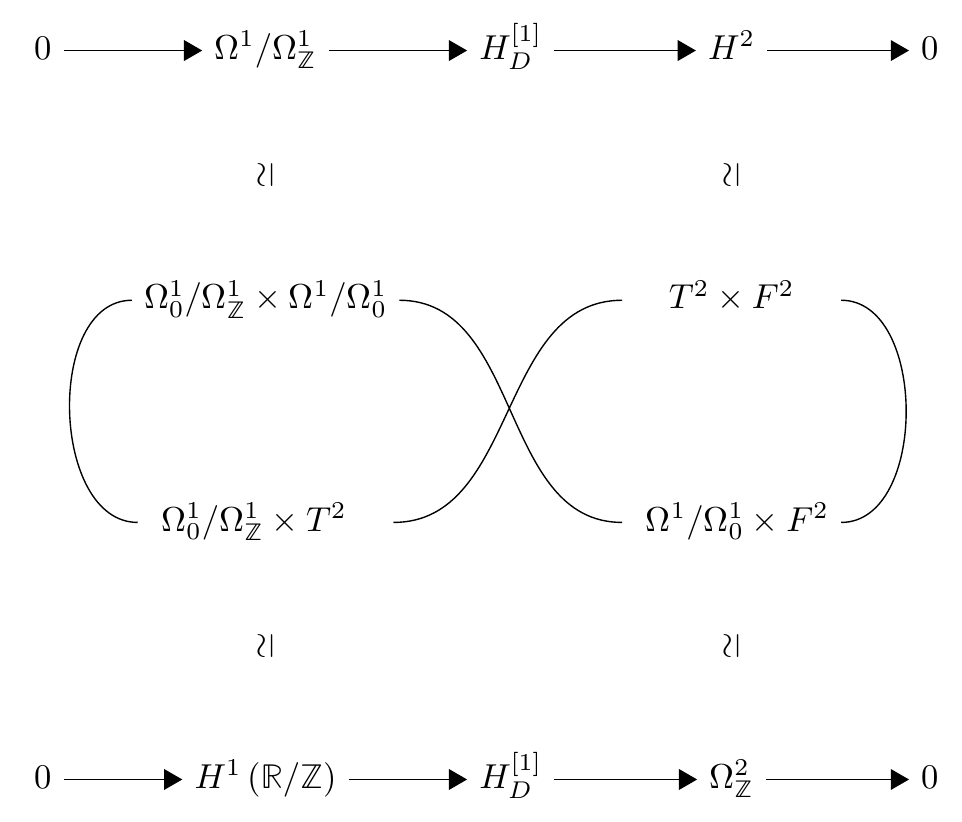}
\end{center}

\section{Operations and duality on Deligne cohomology classes}

Given two Deligne cohomology classes $A$ and $B$ with respective representatives $\left(A_{\alpha},\Lambda_{\alpha\beta},n_{\alpha\beta\gamma}\right)$ and $\left(B_{\alpha},\Theta_{\alpha\beta},m_{\alpha\beta\gamma}\right)$, we define a Deligne cohomology classe $A\star B$ with representative:
\begin{equation}
\left(A_{\alpha}\wedge d B_{\alpha}, \Lambda_{\alpha\beta}B_{\beta}, n_{\alpha\beta\gamma}B_{\gamma},n_{\alpha\beta\gamma}\Theta_{\gamma\rho},n_{\alpha\beta\gamma}m_{\gamma\rho\sigma} \right)
\end{equation}

The integral of a Deligne cohomology class $A$ with representative $\left(A_{\alpha},\Lambda_{\alpha\beta},n_{\alpha\beta\gamma}\right)$ over a cycle $z$ is defined by:
\begin{equation}
\oint_{z}A \underset{\Z}{=} \sum\limits_{\alpha}\int_{z_{\alpha}=U_{\alpha}\cap z} A_{\alpha} - \sum\limits_{\alpha\beta}\int_{z_{\alpha\beta} = U_{\alpha\beta}\cap z}\Lambda_{\alpha\beta}
\end{equation}
where the $\underset{\Z}{=}$ means that the equality is satisfied in $\slfrac{\R}{\Z}$, that is, up to an integer. This integral is nothing but a holonomy, that is, a typical observable of Chern-Simons and BF quantum field theories. This definition ensures gauge invariance in the sense described in the introduction.

We can define in the same way the integral over $M$ of $A\star B$ which provides a generalization of Chern-Simons and BF abelian actions:
\begin{align}
\nonumber\int_{M}A\star B \underset{\Z}{=}\sum\limits_{\alpha}&\int_{U_{\alpha}}A_{\alpha}\wedge d B_{\alpha} - \sum\limits_{\alpha\beta}\int_{U_{\alpha\beta}}\Lambda_{\alpha\beta}B_{\beta} \\
&+ \sum\limits_{\alpha\beta\gamma}\int_{U_{\alpha\beta\gamma}} n_{\alpha\beta\gamma}B_{\gamma} - \sum\limits_{\alpha\beta\gamma\delta}\int_{U_{\alpha\beta\gamma\delta}} n_{\alpha\beta\gamma}\Theta_{\gamma\rho}
\end{align}
Let us point out that the first term is nothing but the local classical action, the other terms ensuring the gluing of local expressions up to an integer.

Note that:
\begin{equation}
\left|
	\begin{array}{ccc}
		Z_{1} \times H^{\left[1\right]}_{D} & \longrightarrow & \slfrac{\R}{\Z} \\
		\left(z , A\right) & \longmapsto & \oint_{z}A \\
	\end{array}
\right. \label{Z1}
\end{equation}
defines a bilinear pairing from the space $Z_{1}$ of $1$-cycles and the space of Deligne cohomology classes (both considered as $\Z$-moduli) in $\slfrac{\R}{\Z}$ as well as:  
\begin{equation}
\left|
	\begin{array}{ccc}
		H^{\left[1\right]}_{D}\times H^{\left[1\right]}_{D} & \longrightarrow & \slfrac{\R}{\Z} \\
		\left(A,B\right) & \longmapsto & \int_{M} A \star B \\
	\end{array}
\right.
\end{equation} 

Starting from that remark, and for later convenience, we will consider Pontrjagin dual $\left(X\right)^{\#} = \mathrm{Hom}\left(X,\slfrac{\R}{\Z}\right)$ of a group $X$. Considering $\mathrm{Hom}$ as a functor, we can show that the following sequences are exact:
\begin{center}
\includegraphics[scale=1.]{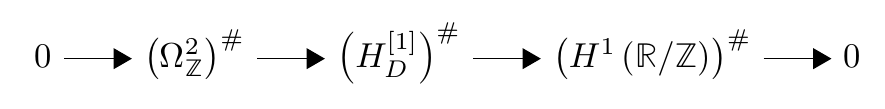}
\end{center}
and:
\begin{center}
\includegraphics[scale=1.]{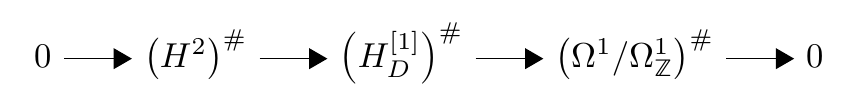}
\end{center}
Moreover, the information of the first two exact sequences is included in those two new ones:
\begin{center}
\includegraphics[scale=1.]{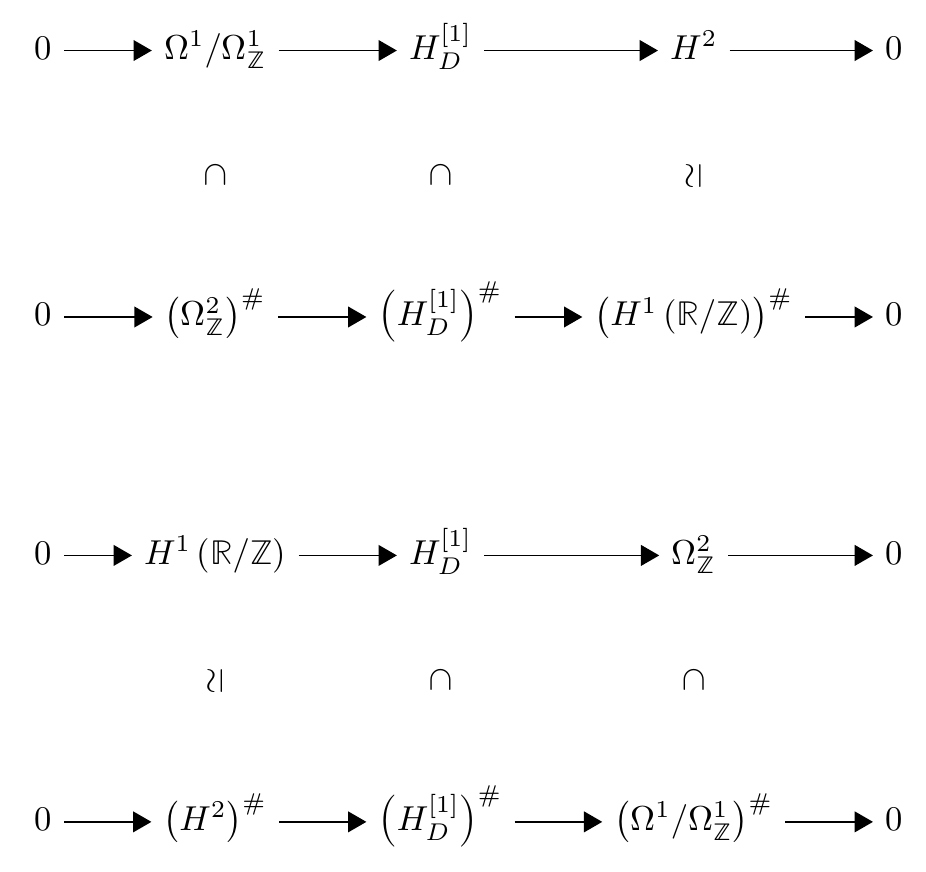}
\end{center}
The Pontrjagin dual is a generalization to distributional objects. Finally, we see that $Z_{1}\subset \left(H^{\left[1\right]}_{D}\right)^{\#}$ in the sense of \eqref{Z1}.  

\section{Decomposition of Deligne cohomology classes}

The structure of Deligne cohomology classes is such that each class $A$ can be decomposed as the sum of an origin indexed on the cohomology of $M$ (basis of the discrete fibre bundle of Deligne cohomology classes) and a translation taken in $\Omega^{1}/\Omega_{\Z}^{1}$:
\begin{equation}
A=A^{0}_{\textbf{a}}+\omega, \,\, \textbf{a}\in H^{2}\left(\Z\right)
\end{equation}
The result of functional integrals over the space of Deligne cohomology classes will not depend on the choice of the origins, but the complexity of the computations will. Thus, our goal is to find convenient origins with algebraic properties that will enable to perform computations easily.

Concerning the translations, we can decompose (non canonically) $\Omega^{1}/\Omega_{\Z}^{1}$ as:
\begin{equation}
\Omega^{1}/\Omega_{\Z}^{1} \simeq \Omega^{1}/\Omega_{0}^{1} \times \Omega^{1}_{0}/\Omega_{\Z}^{1}
\end{equation}
where $\Omega_{0}^{1}$ denotes the set of closed $1$-forms. Furthermore:
\begin{equation}
\Omega^{1}_{0}/\Omega_{\Z}^{1} \simeq \left(\slfrac{\R}{\Z}\right)^{b_{1}}
\end{equation}
$b_{1}$ being the first Betti number. We will call \textbf{zero modes} the elements $\omega_{0}\in \Omega^{1}_{0}/\Omega_{\Z}^{1}$. With this decomposition, we obtain:
\begin{equation}
\forall\,\omega_{0}\in\Omega^{1}_{0}/\Omega_{\Z}^{1},\,\forall\omega\in\Omega^{1}/\Omega_{\Z}^{1},\, \int_{M}\omega\star\omega_{0}\underset{\Z}{=}0
\end{equation}

Let us consider generators $z_{a}$ of the free part of the homology of $M$. Then, by Pontrjagin duality, we can associate to it a unique element $\eta_{z_{a}}\in \left(H^{\left[1\right]}_{D}\right)^{\#}$. Thus, for a fibre over $\sum\limits_{a}m^{a}z_{a}\in F_{1}\simeq F^{2}$ we will consider as origin the element:
\begin{equation}
A_{\textbf{m}}=\sum\limits_{a}m^{a}\eta_{z_{a}}\in \left(H^{\left[1\right]}_{D}\right)^{\#} 
\end{equation}
Note that: 
\begin{equation}
\int_{M}A_{\textbf{m}}\star A_{\textbf{n}}\underset{\Z}{=}0 
\end{equation}
since it represents a linking number which is necessarily an integer. We impose as a convention the so-called \textbf{zero regularisation}:
\begin{equation}
\int_{M}A_{\textbf{m}}\star A_{\textbf{m}}\underset{\Z}{=}0 
\end{equation}
which is ill-defined as a self-linking. Finally, if we decompose $\omega_{0}\in\Omega^{1}_{0}/\Omega_{\Z}^{1}$ as $\omega_{0}=\sum\limits_{b}\theta_{b}\rho^{b}$ with $\oint_{z_{a}}\rho^{b}=\delta_{a}^{b}$, then we obtain:
\begin{equation}
\int_{M}A_{\textbf{m}}\star \omega_{0}\underset{\Z}{=}\textbf{m}\cdot\boldsymbol{\theta} 
\end{equation}

Let us consider now a generator $\tau_{a}$ of the component $\Z_{p_{a}}$ of the torsion part of the homology of $M$. This means that $\tau_{a}$ is the boundary of no surface, but $p_{a}\tau_{a}$ is. Consider now $\eta_{\tau_{a}}\in \left(H^{\left[1\right]}_{D}\right)^{\#}$ defined by:
\begin{equation}
\eta_{\tau_{a}}=\left(0,\frac{m_{\alpha\beta}}{p_{a}},n_{\alpha\beta\gamma}\right)
\end{equation}
where $p_{a}n_{\alpha\beta\gamma} = m_{\beta\gamma}-m_{\alpha\gamma}+m_{\alpha\beta}$. Thus, for a fibre over $\sum\limits_{a}\kappa^{a}\tau_{a}\in T_{1}\simeq T^{2}$ we will consider as origin the element:
\begin{equation}
A^{0}_{\boldsymbol{\kappa}}=\sum\limits_{a}\kappa^{a}\eta_{\tau_{a}}\in \left(H^{\left[1\right]}_{D}\right)^{\#} 
\end{equation}
This choice has several advantages since we can show that:
\begin{equation}
\int_{M}A^{0}_{\boldsymbol{\kappa_{1}}} \star A^{0}_{\boldsymbol{\kappa_{2}}}
\underset{\Z}{=}-Q\left(\boldsymbol{\kappa_{1}},\boldsymbol{\kappa_{2}}\right)
\end{equation}
where $Q$ is the so-called linking form, which is a quadratic form over the torsion of the cohomology. Also:
\begin{equation}
\int_{M}A^{0}_{\boldsymbol{\kappa}} \star A_{\textbf{m}}
\underset{\Z}{=}0
\end{equation}
for any free origin $A_{\textbf{m}}$ and:
\begin{equation}
\int_{M}A^{0}_{\boldsymbol{\kappa}} \star \omega
\underset{\Z}{=}0
\end{equation}
for any translation $\omega$.

\section{$U\left(1\right)$ Chern-Simons and BF theories}

Chern-Simons abelian action is generalized as:
\begin{equation}
S_{\mathrm{CS}_{N}}\left[A\right] = 2\pi N\int_{M}A\star A
\end{equation}
Since $\int_{M}A\star A \in \slfrac{\R}{\Z}$, then $N$ has here to be quantized:
\begin{equation}
N\in\Z.
\end{equation}
The partition function is defined as:
\begin{equation}
Z_{\mathrm{CS}_{N}}=\frac{1}{\mathcal{N}_{\mathrm{CS}_{N}}} \displaystyle\int_{\left(H^{\left[1\right]}_{D}\right)^{\#}} e^{i S_{\tiny\textsc{CS}_{N}}\left[A\right]}\mathcal{D}\!A
\end{equation}
$\mathcal{N}_{\mathrm{CS}_{N}}$ being a normalization that has to cancel the intrinsic divergency of the functional integral. The functional measure we use is then:
\begin{equation}
d\mu_{\mathrm{CS}_{N}}\left[A\right]=\mathcal{D}A e^{iS_{\mathrm{CS}_{N}}\left[A\right]}
\end{equation}
Assume that this measure verifies the so-called Cameron-Martin property, that is:
\begin{equation}
d\mu_{\mathrm{CS}_{N}}\left[A+\omega\right]=d\mu_{\mathrm{CS}_{N}}\left[A\right]e^{4i\pi N\displaystyle\int_{M}A\star\omega}e^{2i\pi N\displaystyle\int_{M}\omega\star\omega}
\end{equation}
for a fixed connection $A$ and a translation $\omega$, then, for $j_{\gamma}$ a translation in $\left(H^{\left[1\right]}_{D}\right)^{\#}$ associated to a cycle $\gamma$:
\begin{equation}
d\mu_{\mathrm{CS}_{N}}\left[A+m\frac{j_{\gamma}}{2N}\right]=d\mu_{\mathrm{CS}_{N}}\left[A\right]
\end{equation}
Using the algebraic properties given before, we can compute exactly the Chern-Simons abelian partition function.

As a convention, for the normalization we choose:
\begin{equation}
\label{Normalization}
\mathcal{N}_{\mathrm{CS}_{N}}=\displaystyle\int_{\left(\slfrac{\Omega^{1}}{\Omega^{1}_{\Z}} \right)^{\#}} e^{i S_{\tiny\textsc{CS}_{N}}\left[\omega\right]}\mathcal{D}\omega=\displaystyle\int_{\left(\slfrac{\Omega^{1}}{\Omega^{1}_{0}} \right)^{\#}} e^{i S_{\tiny\textsc{CS}_{N}}\left[\omega\right]}\mathcal{D}\omega
\end{equation}
which corresponds to the trivial fibre of Deligne bundle for our theory defined over a manifold $M$. This trivial fibre is the (only) one that constitutes Deligne bundle if we consider a theory over $S^{3}$. This choice enables to establish a link with Reshetihin-Turaev abelian invariant (see \cite{MT2015}). Note that usually the normalization of Reshetikhin-Turaev invariant is chosen to be related to $S^{1}\times S^{2}$. However if the normalization is done with respect to $S^{3}$ then one recovers in the abelian case the invariants obtained with convention \eqref{Normalization}.

This way, we find:
\begin{equation}
Z_{\mathrm{CS}_{N}} 
= \displaystyle\sum_{\boldsymbol{\tau_{A}}\in T^{2}} 
e^{-2i\pi N Q\left(\boldsymbol{\tau_{A}},\boldsymbol{\tau_{A}}\right)}
\end{equation}
Analogous considerations apply to BF abelian theory whose generalized action is:
\begin{equation}
S_{\mathrm{BF}_{N}}\left[A,B\right] = 2\pi N\int_{M}A\star B
\end{equation}
($N$ being here also quantized) which leads to a partition function written as:
\begin{equation}
Z_{\mathrm{BF}_{N}} 
= \displaystyle\sum_{\boldsymbol{\tau_{A}}\in T^{2}}\displaystyle\sum_{\boldsymbol{\tau_{B}}\in T^{2}} 
e^{-2i\pi N Q\left(\boldsymbol{\tau_{A}},\boldsymbol{\tau_{B}}\right)}
= \prod\limits_{i=1}^{n}\gcd\left(p_{i},N\right)p_{i}
\end{equation}

Computations of expectation values of observables can also be performed thanks to this method in both Chern-Simons and BF abelian theories (see \cite{MT2015} and \cite{MT2016}).

\section{Conclusion}

Several correspondances in the non-abelian case, mainly $SU\left(2\right)$, have been established formally, that is, with manipulations of ill-defined quantities:
\begin{enumerate}
\item Chern-Simons partition function is related to Reshetikhin-Turaev topological invariant \cite{W1989},
\item BF partition function is related to Turaev-Viro topological invariant \cite{PR1968},
\item the square modulus of Chern-Simons partition function is equal to the BF partition function \cite{CCRFM1995}.
\end{enumerate}
This is summed up on the following diagram:
	\begin{eqnarray*}
		\abs{Z_{\mathrm{CS}_{N}}}^{2} & \overset{\mathrm{Cattaneo}}{\cdots} & Z_{\mathrm{BF}_{N}}\\
		\text{\footnotesize{Witten}\hspace{0.25cm}}\vdots\hspace{0.5cm} &  & \hspace{0.25cm} \vdots \hspace{0.25cm}\text{\footnotesize{Ponzano, Regge... }} \\
		\abs{\mathrm{RT}_{N}}^{2} & \underset{\mathrm{Turaev}}{=} & \mathrm{TV}_{N}
	\end{eqnarray*}
the only result perfectly rigorously established being the one of Turaev, Reshetikhin and Viro (see \cite{RT1990}, \cite{TV1992} and \cite{T2010}).

In the abelian case, we saw that Deligne cohomology approach enables to define rigorously functional integration in the specific case of Chern-Simons and BF theories. Using this tool, we show that the previous diagram is no longer correct and has to be replaced by the following one:
	\begin{eqnarray*}
		\abs{Z_{\mathrm{CS}_{N}}}^{2} & \overset{\mathrm{Thuillier \, \& \, M.}}{\ne} & Z_{\mathrm{BF}_{N}}\\
		\text{\footnotesize{Guadagnini \& Thuillier}\hspace{0.25cm}}\rotatebox{90}{=} \hspace{0.5cm} & & \hspace{0.25cm} \rotatebox{90}{=} \hspace{0.25cm} \text{\footnotesize{Thuillier \& M.}} \\
		\abs{\mathrm{RT}_{N}}^{2} & \ne & \mathrm{TV}_{N}
	\end{eqnarray*}
where the hypothesis of Turaev are not necessarily satisfied with abelian representations, leading to an inequality in general.

This shows that the abelian theories, contrary to what could be expected, are not a simple trivial subcase of the non-abelian ones. However, we expect to find some traces of this abelian case in the non-abelian one, which is the aim of present works.

\vspace{1.cm}

\textit{The author declares that there is no conflict of interest regarding the publication of this paper.}

\end{document}